\documentclass[superscriptaddress,twocolumn,showpacs,showkeys,amsmath,prb,final]{revtex4}

\usepackage{graphicx}
\usepackage{dcolumn}
\usepackage[sort&compress]{natbib}
\usepackage{subfigure}
\usepackage{ifpdf}

\ifpdf
\usepackage{hyperref}
\hypersetup{%
pdfauthor={Mauro M. Doria, Antonio R. de C. Romaguera and F. M.
Peeters},
pdftitle={Effect of the boundary condition on the vortex patterns in mesoscopic three-dimensional
superconductors - disk and sphere},
pdfsubject={Vortex theory},
pdfkeywords={Department of Physics , Universiteit Antwerpen
 Groenenborgerlaan 171 / U 208 B-2020 Antwerpen},
baseurl={http://www.cmt.ua.ac.be} }
\fi

\begin{document}
\title{Effect of the boundary condition on the vortex patterns in mesoscopic three-dimensional superconductors - disk and sphere}
\author{Mauro M. Doria}%
\email{mmd@if.ufrj.br}\homepage{http://www.if.ufrj.br/~mmd}%
\affiliation{Instituto de F\'{\i}sica, Universidade Federal do Rio de
Janeiro, C. P. 68528, 21941-972, Rio de Janeiro, Brazil}%
\affiliation{Departement Fysica, Universiteit Antwerpen, Groenenborgerlaan 171, B-2020
Antwerpen, Belgium}%
\author{Antonio R. de C. Romaguera}%
\affiliation{Instituto de F\'{\i}sica, Universidade Federal do Rio de
Janeiro, C. P. 68528, 21941-972, Rio de Janeiro, Brazil}%
\affiliation{Departement Fysica, Universiteit Antwerpen, Groenenborgerlaan 171, B-2020
Antwerpen, Belgium}%
\author{F. M. Peeters}%
\email{francois.peeters@ua.ac.be}\homepage{http://www.cmt.ua.ac.be} \affiliation{Departement Fysica,
Universiteit Antwerpen, Groenenborgerlaan 171, B-2020 Antwerpen, Belgium}

\date{\today}

\begin{abstract}
The vortex state of mesoscopic three-dimensional superconductors is determined using a minimization procedure of
the Ginzburg-Landau free energy. We obtain the vortex pattern for a mesoscopic superconducting sphere and find
that vortex lines are naturally bent and are closest to each other at the equatorial plane. For a
superconducting disk with finite height, and under an applied magnetic field perpendicular to its major surface,
we find that our method gives results consistent with previous calculations. The matching fields, the
magnetization and $H_{c3}$, are obtained for models that differ according to their boundary properties. A change
of the Ginzburg-Landau parameters near the surface can substantially enhance $H_{c3}$ as shown here.
\end{abstract}
\preprint{APS/123-QED}
\maketitle

\section{\label{sec:intro}Introduction}

In the last decade the response of a mesoscopic superconducting disk to a perpendicular magnetic field has been
theoretically\cite{PhysRevLett.81.2783,PhysRevB.57.13817} and experimentally\cite{Nature.390.259} studied. The
small volume to surface ratio of mesoscopic superconductors brings new and interesting physical properties such
as giant vortices, recently detected thanks to new advances in small-tunnel-junction
technology\cite{PhysRevLett.93.257002}. Previous studies of mesoscopic systems were based on two dimensional
(2D) theory, where the superconducting condensate was assumed not to vary along the direction of the magnetic
field. This assumption is not taken here and we study finite size extreme type-II mesoscopic superconductors
using a truly three-dimensional (3D) theoretical approach previously applied to a bulk
superconductor\cite{EuroPhysLett.67.446,PhysRevB.66.064519}.

Only a few vortices fit inside a mesoscopic superconductor whereas
for a bulk superconductor, with non-superconducting inclusions
inside, the number of vortices is uncountable. By \textit{inclusion}
we refer to a pinning center with the size equal to a few multiples
of the coherence length, $\xi $. These two systems have a similar
properties because of their mesoscopic scale structure. For
instance, giant vortices are naturally found in mesoscopic
superconductors but not in bulk superconductors, where the
nucleation of a vortex line with multiple magnetic flux N$\Phi_0$ is
energetically forbidden and only the nucleation of N individual
vortex lines with $\Phi_0$ is possible ($\Phi_0$ is the quantum of
flux). However this picture does not hold in the presence of
inclusions. For instance for a bulk superconductor, Mkrtchyan and
Shmidt\cite{SovietPhys.JETP.34}, Buzdin\cite{PhysRevB.47.11416}, and
some of us\cite{PhysRevB.60.13164,PhysRevLett.96.207001} have shown
that a columnar defect can hold a multiple magnetic flux N$\Phi_0$.

Metastability, matching fields, occupation numbers and giant
vortices have been experimentally studied in 2D bulk superconductors
with inclusions, namely, superconducting films with an array of
two-dimensional mesoscopic pinning centers consisting of not fully
perforated holes (blind holes)\cite{PhysRevB.70.24509}, fully
perforated holes (open
holes)\cite{PhysRevB.51.3718,PhysLettA.195.373} and micro
holes\cite{PhysRevB.54.7385,PhysRevB.70.54515}. A similar 3D bulk
superconductor with a truly three-dimensional arrangement of
internal inclusions has yet to be experimentally realized though it
has been theoretically
studied\cite{PhysRevB.60.13164,PhysRevLett.96.207001}. Such internal
inclusions are present as a random
array\cite{sup.sci.tech.9.1015,sup.sci.tech.18.l13} in the
LRE–Ba–Cu–O superconductors (where LRE is a light rare earth element
such as Nd, Sm, Eu and Gd) and some of the present ideas may be
useful to explain their unusual properties. Such internal inclusions
bring new features to vortex physics as just a single one can trap
many vortex lines in its neighborhood. The regular array theoretical
study of the 3D bulk superconductor with inclusions was done in the
context of a modified version of the Ginzburg-Landau (GL) theory.
Other studies based on the Ginzburg-Landau theory for 3D systems
have been done, including shells \cite{j.math.phts.46.095109} and
constricted superconducting wires\cite{europhyslett.74.151} both in
the extreme type-II limit.

In this paper we apply the same theoretical approach to study a
mesoscopic superconductor. A 3D lattice of inclusions turns into a
3D lattice of mesoscopic superconductors with the same geometry when
the insulating regions are replaced by superconducting ones and
vice-versa. The 3D lattice of mesoscopic superconductors becomes a
set of individually equivalent mesoscopic superconductors for a
London penetration length much larger than their size. In this case
the local field is constant and equal to the applied field
everywhere. In this way we obtain the vortex patterns for a single
mesoscopic superconductor, here of a sphere and of a disk. In the
past the vortex patterns of mesoscopic superconductors were obtained
\cite{PhysRevB.70.024508,PhysRevB.65.104515,PhysRevLett.81.2783,PhysRevB.57.13817}
in the limit of an extremely thin film. This condition renders the
variation of the Cooper pair density along the magnetic field
negligible. Baelus and Peeters\cite{PhysRevB.65.104515} studied
several different flat geometries typically with thickness $0.1\xi$
and obtained the vortex patterns from the two-dimensional GL
equations supplemented by the Saint-James--de
Gennes\cite{Phys.Lett.7.306} boundary conditions at the edge. They
considered a Ginzburg-Landau parameter, the ratio of the London
penetration to the coherence length, $\kappa=0.28$, and solved the
two GL equations. Here we study a thick disk and compare our results
to theirs\cite{PhysRevB.65.104515} taking into account that in their
case the magnetic response to an external applied field is much
stronger than here. We only report results for $\kappa \rightarrow
\infty$ although our method is not restricted to this limit.  Notice
that for our case the boundary conditions are truly
three-dimensional, and so, imposed in all directions including
perpendicularly to the flat geometry.

The major new results of this paper can be summarized as follows.
(i) We find the vortex pattern for a mesoscopic sphere, with radius
$R_s=4.0\xi$, a problem whose solution is beyond the scope of
previous 2D techniques. (ii) We show that a slight change of the
Ginzburg-Landau parameters near the edge can substantially increase
the $H_{c3}$ field. A thin layer covers the superconductor and
separates it from the outside insulating world. This layer is also
superconducting but with effective GL parameter slightly different
from those inside. For a bulk system the phenomenological GL
parameters are known to be related to the microscopic parameters in
the following way: $\alpha_0 \sim (kT_c)^2/\epsilon_F$, $\beta \sim
(kT_c)^2/(\epsilon_F n)$, where $T_c$ is the critical temperature,
$\epsilon_F$ is the Fermi energy, and $n$ is the electronic density.
Similar relations should exist in case of a mesoscopic
superconductor although we don't obtain them here. We just show that
a decrease near the edges of the effective Cooper pair mass, $m$,
and of $\alpha_0$, lead to an enhancement of $H_{c3}$. Therefore the
present approach is interesting for a system with a small volume to
surface ratio because there a slight change at the boundary over a
distance less or equal to $\xi$ is found here to make a significant
difference. The present approach relies on a free energy
minimization procedure carried in the whole space, including the
world outside the superconductor, where the order parameter is found
to vanish. The decay of the Cooper pair density at the boundary,
from a finite value inside the mesoscopic superconductor to zero
outside, is treated here. Notice that standard differential equation
approaches, such as that of Ref. \onlinecite{PhysRevB.65.104515},
only treat the volume internal to the superconductor, and do not
treat the order parameter discontinuity at the edge, from a finite
value to zero at the outside world.  In this paper we study three
kinds of boundary conditions and discuss them in the context of a
disk of radius $R=4.0\xi$.

Below we provide a short description of the disk and sphere boundary problems treated here. We chose to give
them names that recall their major features: (i) \textit{sharp}: a disk is considered and its boundary treatment
is the standard one used for comparison with all other models. A coarse grained grid is used and gives a fast
and efficient convergence to the final configuration. The vortex states are satisfactorily described here. Its
name stems from the sharp definition of the edge. (ii) \textit{mesh}: this model is the same as \textit{sharp}
except with a refined grid, which contains 8.2 times more grid points. (iii) \textit{sphere}: a sphere is
treated here with the same grid coarseness along the disk radial direction as in the \textit{sharp} model.(iv)
\textit{BP2D}: this is the disk reported by Baelus and Peeters\cite{PhysRevB.65.104515} using their
two-dimensional approach. (v) \textit{smooth}: this type of boundary was previously used in
Refs.~\onlinecite{Brazilian.J.Phys.35,Eur.Phys.J.B.42} for insulating pinning spheres inside a bulk
superconductor and is used here for the disk. Its major property is that the supercurrent normal to the surface
does not disappear abruptly but over some small region (fraction of $\xi$). (vi) \textit{step}: This model
contains a superconducting layer that sets the disappearance of superconductivity. Thus there are two concentric
disks and we find that this intermediate layer stabilizes the superconducting state in the inner disk. This
model features a very high $H_{c3}$ as compared to the other models.

The paper is organized as follows. In  Section~\ref{sec:the} we describe our theoretical approach valid for the
following two complementary situations: (i)superconductor with non-superconducting inclusions and (ii)mesoscopic
superconductor. In Section~\ref{sec:mod} we describe the disk and sphere boundary models and discuss their
properties obtained through our numerical simulations. In Section~\ref{sec:dis} we compare the models and
discuss many of their common features. Finally in Section~\ref{sec:con} we summarize the main achievements of
this work.

\section{Theoretical Aspects}
\label{sec:the}

One of the advantages of the present method is that it can easily incorporate the shape of the mesoscopic
superconductor, which is done at the free energy level, given by the density expansion below,
\begin{eqnarray}
F = \int \frac{dv}{V} &&\Biggl [ \tau(\vec{r})  \frac{|\vec D
\psi|^2}{2 m} + \tau(\vec{r})\alpha_0 (T - T_c) |\psi|^2 +\nonumber\\
&&+ \frac{\beta}{2} |\psi|^4 + \frac{\vec h^2}{8\pi} \Biggl ],
\label{eq:glth}
\end{eqnarray}
where $\vec D \equiv (\hbar/i) \vec \nabla - q \vec A / c $, $q$ is the Cooper pair charge, and $\tau(\vec{r})$
is a step-like function, equal to one inside the mesoscopic superconductor and zero outside. The $\tau(\vec{r})$
contains the geometry of the mesoscopic superconductor. The corresponding Euler-Lagrange equations are given by,

\begin{eqnarray}
&\frac{i\hbar}{2 m}\vec \nabla \tau(\vec{r}) \cdot \vec D \psi& +
\tau(\vec{r}) \frac{{\vec D}^2 \psi}{2 m} +%
\nonumber \\&& %
+\tau(\vec{r})\alpha_0 (T - T_c) \psi +  \beta |\psi|^2 \psi =0, \label{eq:gleq1}
\end{eqnarray}
\begin{eqnarray}
&&\vec \nabla \times \vec h = \frac{4\pi \vec J }{c},%
\vec J = \frac{q}{2m}\tau(\vec{r}) \Big [ \psi^{*} \vec D \psi + (\vec D \psi)^{*}\psi \Big ]. \label{eq:gleq2}
\end{eqnarray}
These modified GL equations automatically incorporate the appropriate boundary conditions through the step-like
function $\tau(\vec{r})$, discontinuous at the edge, equal to one inside, and zero outside, for the mesoscopic
superconductor. The gradient of $\tau (\vec{r})$ is zero everywhere except at the interface, where it diverges.
Any finite and physical solution must obey $\vec \nabla \tau \cdot \vec D \psi = 0 $ because this divergence is
proportional to the Dirac delta function. For instance, along the radial direction of the disk: $\tau(r) = 1$
for $r \le R$ and $\tau(r) = 0$ for $x > R$, thus the derivative becomes $\vec \nabla \tau = \hat r
\partial \tau(r)/\partial r = - \hat r \delta(r-R)$. Let $f(r)$ be
any function describing products of the order parameter and its derivatives. The product $f(r)\partial
\tau(r)/\partial r$ diverges at the border and the only way to make it vanish there is through the condition
$f(R)=0$. Thus the well-known Saint-James--de Gennes\cite{Phys.Lett.7.306} boundary condition, $ \hat n \cdot
\vec D \psi\big |_n = 0$, is recovered here. As the superconducting order parameter is defined everywhere in the
unit cell, including \textit{outside} the mesoscopic superconductor, where Eq.~(\ref{eq:glth}) becomes $F =
(1/V) \int dv \lbrack \; \beta |\psi|^4/2 + \vec h^2/8\pi\; \rbrack$ outside the superconducting volume. It must
vanish as a result of the free energy minimization, and variation with respect to $\psi$ and $\vec A$ shows that
the minimum is reached for $\psi = 0$ and $\vec \nabla \times \vec h = 0$ according to
Eqs.~(\ref{eq:gleq1})-(\ref{eq:gleq2}). It is possible to obtain more elaborate versions of the GL theory, such
as the one containing a local depression of the critical temperature through a function $T_c(\vec{r})$
\cite{PhysRevB.60.13164} in Eq.~(\ref{eq:gleq1}). In this work the free energy is normalized by the constant
$F_0=H_c^2/8\pi$ and all fields are normalized in terms of the upper critical field, $H_{c2}$. Lengths are in
units of the coherence length, $\xi(T) = \sqrt{\hbar^2/2m\alpha_0(T_c-T)}$, and the density $|\psi|^2$ is
normalized by $(\alpha_0(T_c-T)/\beta)$, such that its maximum value of 1 is reached, for instance, inside a
bulk superconductor (no boundaries) for zero applied field.

We stress some differences in the application of the present method to the two complementary problems. For the
former case the magnetic induction, $\vec B = \int dv \; \vec h /V$, is constant whereas for the latter the
applied field $ \vec H$ is constant. In the former case the unit cell edges are fully inside the superconductor
and this introduces into the theory integers associated to the periodic boundary conditions imposed by the unit
cell. These integers follow the condition that the order parameter be single-valued. Though $|\psi|^2$ and $\vec
h$ are periodic, $\psi$ and $\vec A$ only need to coincide at the unit cell surfaces up to a gauge
transformation, whose expression gives room to introduce these integers.
\begin{eqnarray}
\psi(\vec{r}+\vec{L}_\mu)&=&e^{\frac{i2\pi}{\phi_0}\Lambda_\mu(\vec{r})}\psi(\vec{r})\\
\vec{A}(\vec{r}+\vec{L}_\mu)&=&\vec{A}(\vec{r})+\nabla
\Lambda_\mu(\vec{r})
\end{eqnarray}
where $L_\mu$ is the unit cell length, $\Lambda_\mu(\vec{r})$ is a scalar gauge function and  $\mu=$ $x$, $y$ or
$z$. The minimization procedure shows that such integers are nothing but the number of vortices in the unit
cell, and the magnetic induction is fully determined by them. However for the complementary problem, the
mesoscopic superconductor is fully inside the unit cell and its boundaries are away from the unit cell edges.
Consequently there is no single-valued condition on the order parameter and so, these integers do not exist at
all. Consequently, the independent thermodynamic field in this case, which is the applied field $\vec H$, varies
continuously.

Since there are no screening currents the local field, defined as
$\vec h  = \vec \nabla \times \vec A$, is the external applied field
$\vec h = \vec H$. In this large $\kappa$ limit the magnetization is
directly determined from $\vec M = const \int dv \; \vec r \times
\vec J$, where $\vec J$ is the supercurrent. An extra condition
determines the remaining free parameter $const$, and consequently
the demagnetization constant $D$ of the mesoscopic superconductor:
for small $\vec H$, that is, in the Meissner phase, we impose the
condition that $\vec H +4\pi D \vec M=0$. In contrast, in the
approach of Baelus and Peeters\cite{PhysRevB.65.104515} for finite
$\kappa$, the magnetization is directly obtained from the difference
between the magnetic induction and the applied field.

The minimization of the GL free energy, done numerically through the
so-called Simulated Annealing
method\cite{PhysRevB.41.6335,PhysRevB.39.9573}, is carried in a
discrete three-dimensional space. The free energy is adapted to keep
its gauge invariance in this discrete space. A cell, that consists
of an orthorhombic box containing $N_x.N_y.N_z$ points for this
purpose. Every point in this cell, belonging to the mesoscopic
superconductor or not, has associated to it the fields
$\psi(n_x,n_y,n_z)$, and $\vec A(n_x,n_y,n_z)$, where $n_x =
1,\ldots, N_x$, $n_y=1,\ldots, N_y$, and $n_z=1,\ldots, N_z$. The
physical volume of the box is $L_x\cdot L_y\cdot L_z$, where
$L_x=a_x(N_x-1)$, $L_y=a_y(N_y-1)$ and $L_z=a_z(N_z-1)$. The
distance between two consecutive points along the axes of the box is
$a_x$, $a_y$, and $a_z$. The discrete theory, given by
Eq.~(\ref{eq:dsgl}), properly describes the properties of the
continuous theory under the condition that $\xi$ be much larger than
$a_x$, $a_y$, and $a_z$, the grid resolution.

In the discrete free energy, given by Eq.~(\ref{eq:dsgl}), grid points outside and inside the mesoscopic
superconductor are coupled through gradient terms. For instance in case of no applied field, this coupling is
proportional to $|\psi'_{out}-\psi_{in} |^2$. The fact that $\psi'_{out} \rightarrow 0$ causes $\psi_{in}$ to
get a lower value than deep inside the sample, where the kinetic energy vanishes as the order parameter is the
same in all points. For this reason the density $|\psi|^2$ never reaches its maximum bulk value due to the small
volume to surface ratio.
\begin{widetext}
\begin{eqnarray}
F &=&  \frac{1}{N_x N_y N_z}\sum_{n_x=1}^{N_x}
\sum_{n_y=1}^{N_y}\sum_{n_z=1}^{N_z}\Biggl \{
\frac{\hbar^2}{2m}\frac{1}{a_x^2} \;\frac{\tau(\vec n+\hat
x)+\tau(\vec n)}{2} |\psi(\vec n+\hat
x)-e^{i\frac{2\pi a_x}{\Phi_0}A_x(\vec n)}\psi(\vec n)|^2 + \nonumber \\
&+& \frac{\hbar^2}{2m}\frac{1}{a_y^2} \;\frac{\tau(\vec n+\hat
y)+\tau(\vec n)}{2} |\psi(\vec n+\hat y)-e^{i\frac{2\pi
a_y}{\Phi_0}A_y(\vec n)}\psi(\vec n)|^2 +
\frac{\hbar^2}{2m}\frac{1}{a_z^2} \;\frac{\tau(\vec n+\hat
z)+\tau(\vec n)}{2} |\psi(\vec n+\hat z)-e^{i\frac{2\pi
a_z}{\Phi_0}A_z(\vec n)}\psi(\vec n)|^2+ \nonumber \\ &+& \tau(\vec
n)\alpha_0 (T - T_c) |\psi(\vec n)|^2 + \frac{\beta}{2} |\psi(\vec
n)|^4\Biggl \}.\label{eq:dsgl}
\end{eqnarray}
\end{widetext}
\section{Comparison of the different models}
\label{sec:mod}

The models introduced in Section~\ref{sec:intro} have their
properties summarized in Tables~\ref{tab:model} and
~\ref{tab:modpar} and some of their free energy and magnetization
properties are described in Tables~\ref{tab:hmatch} and
~\ref{tab:magmax}, respectively. The function $\tau(\vec{r})$ is
taken as the product of independent orthogonal direction functions
for all models studied here. Below is a summary of their features.
(i) \textit{sharp}: This model treats the boundaries of a disk of
radius $R=4.0\xi$ and height $d=1.0\xi$ through a $\tau(\vec{r})$
function, 1 inside and 0 outside, both along the radial and the
major axis direction: $\tau(\vec{r})=\tau_\rho(\rho)\cdot\tau_d(z)$.
(ii) \textit{mesh}: The same disk and boundaries of the
\textit{sharp} model is treated here but with a denser grid,
$61\times 61\times 26$ instead of $41\times 41\times 7$. (iii)
\textit{sphere} - This model treats a sphere with stepwise
$\tau(\vec{r})$ function such as in the \textit{sharp} model. (iv)
\textit{BP2D}: This is the disk of
Ref.~\onlinecite{PhysRevB.65.104515}. It has a $0.1\xi$ height and a
$128\times 128\times 1$ grid is used. Although of its
two-dimensional treatment it contains $1.4$ times more grid points
than the three-dimensional \textit{sharp} disk model. (v)
\textit{smooth}: For this model a larger height is taken, $2.0\xi$,
to help stabilize the order parameter inside the disk. The
smoothness  of $\tau_d(z)$, which reaches values below 1 inside the
region $|z|\le d/2$), tends to downgrade the order parameter inside
the disk. For our numerical study we selected, for the exponential
parameter of Table~\ref{tab:model}, N=8. (vi) \textit{step}: The
height is $d=1.5\xi$ and the $\tau(\vec{r})$ function varies
stepwise, taking values 0, 0.8 and 1. The choice of intermediate
value 0.8 is rather arbitrary and we have found that lowering this
intermediate value to 0.5, for instance, causes a substantial
increase of $H_{c3}$, as compared to here. Thus a drop of $\tau$
near the border, and so of the corresponding GL parameters, can
severely affect $H_{c3}$.
\begin{table}[b]
\caption{\label{tab:model} The different models considered in this
paper where $\tau(\vec{r}) = \tau_\rho(\rho)\cdot \tau_d(z)$. $R_s$
is the sphere radius. $R_i$ and $d_i$ are the internal disk radius
and height, respectively.}
\begin{ruledtabular}
\begin{tabular}{|l|l|l|}
Model & ~~~~~~~~~$\tau_\rho(\rho)$ & $~~~~~~~~~~\tau_d(z)$ \\
\hline%
\textit{sharp} &$\tau_\rho\!=\!\begin{cases}1& \rho\le R\cr 0&
\rho>R\cr\end{cases} $ &$\tau_d\!=\!\begin{cases} 1& 2|z|\le d\cr
0&2|z|>d \cr\end{cases}$\\
\hline%
\textit{mesh} & idem & idem \\ \hline
\textit{sphere}&$\tau(\vec{r})\!=\!\begin{cases} 1 &r \le R_s \cr 0
& r > R_s \cr
\end{cases}$& - \\ \hline
\textit{BP2D}& dif. eq.& - \\ \hline
\textit{smooth}&$\tau_\rho\!=\!2/[1+e^{(\rho/R)^N}]$&$\tau_d\!=\!2/[1+e^{(2|z|/d)^N}]$\\
\hline
\textit{step}&$\tau_\rho\!=\!\begin{cases}1&\!\!\!\rho\!\le\!R_i\cr
0.8&\!\!\!R_i\!<\!\rho\!\le\!R\cr
0&\!\!\!\rho\!>\!R\cr\end{cases}$&$\tau_d\!=\!\begin{cases}1&\!\!\!2|z|\!\le\!d_i\cr 0.8&\!\!\!d_i\!<\!2|z|\!\le\!d\cr 0&\!\!\!2|z|\!>\!d\cr\end{cases}$ \\
\end{tabular}
\end{ruledtabular}
\end{table}
%
\begin{table}[t]
\caption{\label{tab:modpar} The parameters of the different models
used in our numerical calculation.}
\begin{ruledtabular}
\begin{tabular}{|l|l|l|l|}
model&grid\footnotemark[1]&cell\footnotemark[2]&parameters\footnotemark[3]\\
\hline \textit{sharp}
&(41,41,7)&(0.3,0.3,0.5)&d=1.0 \\ \hline \textit{mesh} & (61,61,26) &(0.2,0.2,0.2)&d=1.0 \\
\hline \textit{sphere}& (41,41,41) &(0.3,0.3,0.3)& R$_s$=4.0 \\ \hline \textit{BP2D} & (128,128,1) & (0,0,-) & \\
\hline
\textit{smooth} & (41,41,13) & (0.3,0.3,0.5) &d=2.0,N=30 \\
\hline
\textit{step}& (41,41,7)  & (0.3,0.3,0.5) &d=1.5,R$_i$=3.5, d$_i$=0.5\\
\end{tabular}
\end{ruledtabular}
\footnotetext[1]{The number of grid points for the three cell
directions: ($N_x,N_y,N_z$). } \footnotetext[2]{The lattice spacing
for the three cell directions: ($a_x,a_y,a_z$).}
\footnotetext[3]{All the lengths are in units of $\xi$. The disk
radius is $R=4.0$ for all cases.}
\end{table}
%
\begin{table}[b]
\caption{\label{tab:hmatch} The matching fields h$_{L\,L+1}$ between
the angular momentum states L and L+1 for the different models
considered here.}
\begin{ruledtabular}
\begin{tabular}{|l|l|l|l|l|l|l|}
h$_{i\,i+1}$&sharp& mesh& sphere&BP2D& smooth&step\\
\hline
h$_{0\,1}$&0.31&0.30&0.41  &0.39&0.30&0.31\\
h$_{1\,2}$&0.51&0.51&0.65&0.59 &0.52&0.54\\
h$_{2\,3}$&0.69&0.68&0.84&0.74 &0.69&0.72\\
h$_{3\,4}$&0.85&0.83&1.00&0.89&0.84&0.88\\
h$_{4\,5}$&0.99&0.98&1.15&1.02&0.99&1.03\\
h$_{5\,6}$&1.14&1.12&1.28&1.16&1.13&1.20\\
h$_{6\,7}$&1.28&1.26&1.42&1.30 &1.27&1.34\\
h$_{7\,8}$&1.43&1.41&1.56&1.43&1.41&1.48\\
h$_{8\,9}$&1.57&1.54& -&1.57&1.54&1.62\\
h$_{9\,10}$&1.70&1.69&-&1.71&1.68&1.84\\
h$_{10\,11}$&1.87&1.82&-&1.84&-&2.05\\
h$_{11\,12}$&-&-&-&-&-&2.19\\
\end{tabular}
\end{ruledtabular}
\end{table}
\begin{figure}[b]
\centering
\includegraphics[width=\linewidth]{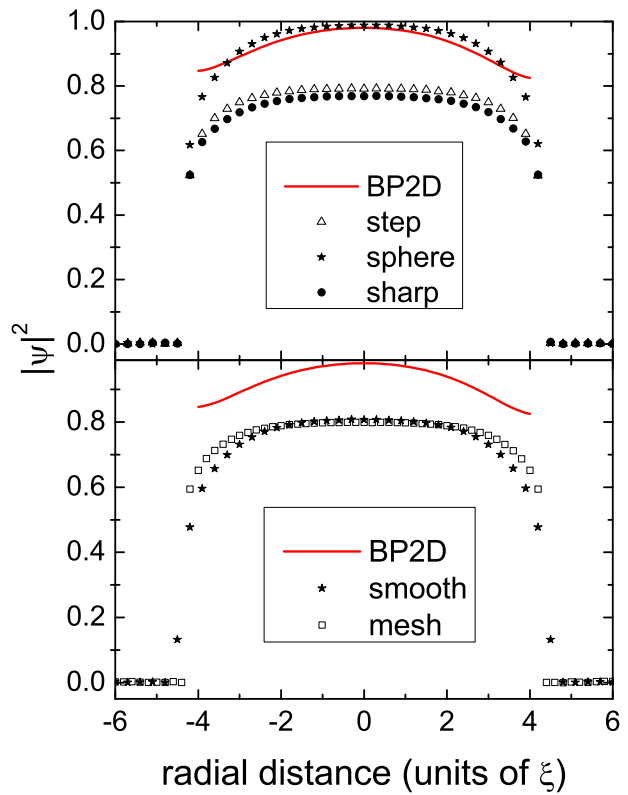}
\caption{(Color online) Cooper pair density $|\Psi|^2$ vs. the distance from the center of the disk for the case
of the Meissner state in the absence of an applied field. The symbols correspond to $|\Psi|^2$ values at the
mesh grid points.} \label{dens1d}
\end{figure}
Fig.~\ref{dens1d} shows the normalized density for zero applied
field (Meissner phase) versus the distance from the geometric center
of the disk along the radial direction, and in case of the sphere,
this distance is along the radial direction inside the equatorial
plane. For clarity the six models considered here were split into
two subsets shown in different plots. For comparison purposes the
\textit{BP2D} model is shown in both plots (red). Notice that for
the \textit{BP2D} model, as well as for the sphere, the maximum
density is 1.0, but not for the other disk models whose maximum
density is about 0.8. The sphere has a larger volume than the disk,
and so, the surface is not so effective to alter the order parameter
in its center.

In presence of an applied field the numerical simulation is carried
in the following way. For zero applied field a random configuration
of the order parameter is assumed inside the cell and a search for
the minimum of the free energy is carried out. The applied field is
increased at constant steps and for a given field one assumes as the
starting order parameter configuration the one found for the
previous field. This procedure is carried sequentially until the
last critical field, $H_{c3}$, is reached and the order parameter
vanishes everywhere. Next the applied field is lowered backwards to
zero applied field. A typical feature of mesoscopic
superconductors\cite{PhysRevLett.81.2783,PhysRevB.57.13817} is a
saw-tooth structure for the descending field of magnetization curve.
The two curves do not coincide, the ascending one has a stronger
diamagnetic signal than the descending curve. The mesoscopic
superconductor exhibits hysteresis as observed, e. g., in Al
disks\cite{Nature.390.259}, and theoretically obtained in previous
studies\cite{PhysRevLett.81.2783,PhysRevB.65.104515}. Metastability
is also found in the complementary problem of  bulk superconductors
with inclusions\cite{EuroPhysLett.67.446,PhysRevB.66.064519}.

\begin{figure}[b]
\centering
\includegraphics[width=1.0\linewidth,height=1.0\linewidth]{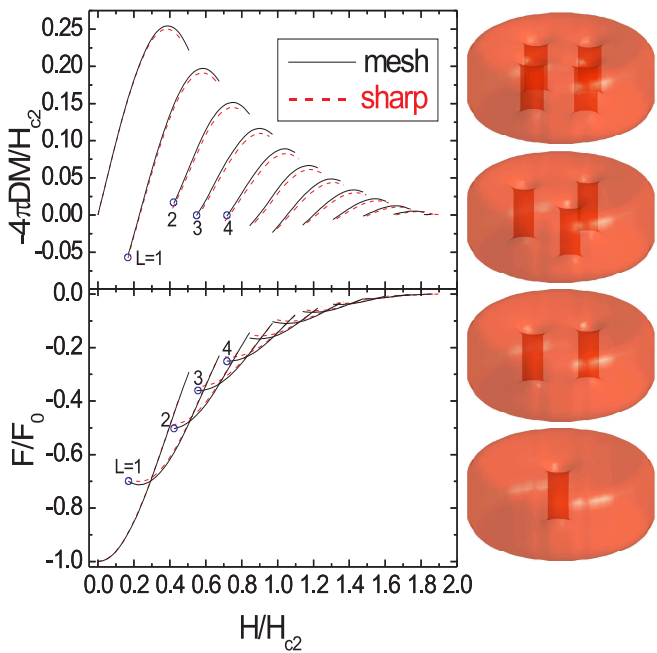}
\caption{(Color online) The \textit{sharp} (red) and the
\textit{mesh} disks free-energy and magnetization curves are shown
here. Iso-density plots for selected applied fields are shown here
to illustrate the first four cases of vortex patterns found for the
thick disk. The three-dimensional figures are iso-contours taken at
20$\%$ of the maximum density $|\psi|^2$. Each iso-contour is a
single surface, the sum of the vortices and the external surface.}
\label{mesh_sharp}
\end{figure}

\begin{figure*}[t]
\centering
\includegraphics[width=\linewidth]{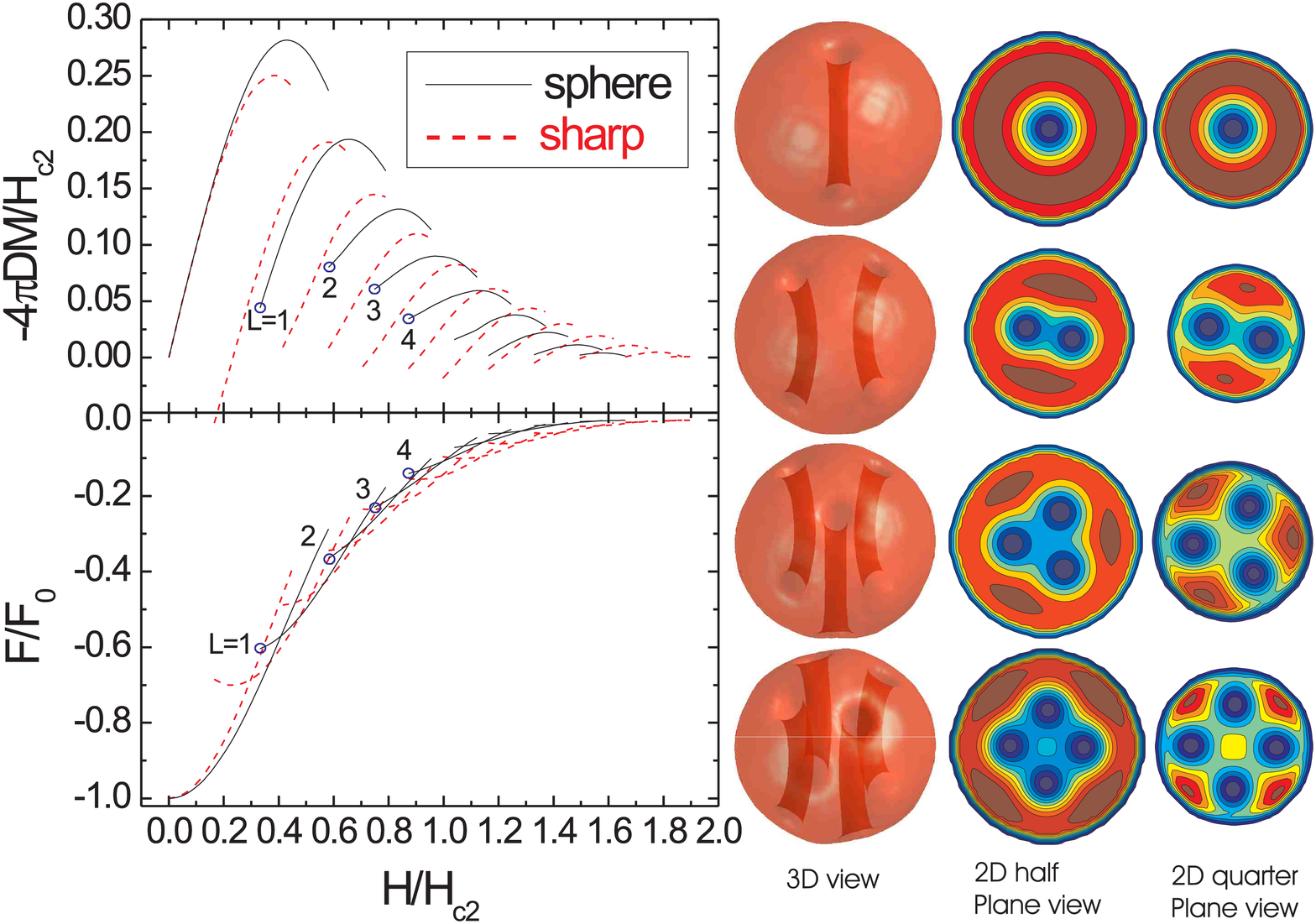}
\caption{(Color online) The \textit{sphere} and the \textit{sharp}
(red) disk free-energy and magnetization curves are shown in the
right panels. Three-dimensional iso-density plots and two
dimensional density contour plots of $|\psi|^2$ are also shown in
the right panels. Each three-dimensional iso-contour is a single
surface, made of the sum of the vortices and the external surface.
Two-dimensional contour plots are taken at the half (equatorial) and
at the quarter plane that cuts the sphere perpendicularly to the
applied field direction.} \label{esf_sharp}
\end{figure*}

Notice that all the magnetization curves shown here
(Figs.~\ref{mesh_sharp},~\ref{esf_sharp},~\ref{fig4}) decompose into
independent non-intersecting lines and result from ancestor curves
that contain their ascending and descending branches. The saw-tooth
structure is a sum of segments, which are parts of the independent
non-intersecting lines. Two consecutive segments are connected by a
vertical jump. The reason for such vertical jumps resides in the
free energy curve which also consists of independent but
intersecting lines. In fact these intersections define the so-called
matching fields which correspond to cross-sections of free energy
lines of neighboring states. Above the matching field the free
energy of the higher state is lower than that of the preceding one
and thus the higher state is preferred. This is also the reason for
the saw-tooth character of the hysteresis curve. Depending on how
the numerical procedure is carried (the magnetic field step, the
simulated annealing temperature, etc) one obtains a different
saw-tooth structure that always falls over the same set of
independent non-intersecting lines. The presence of distinct lines
in the magnetization and free energy curves reveals a parameter that
remains constant upon field sweep. A look at the order parameter
phase reveals that it takes variations from $0$ to $2\pi$ and the
number of such variations remains constant throughout a
magnetization line. Thus this parameter is the total angular
momentum $L$\cite{PhysRevB.57.13817,PhysRevLett.81.2783}. Along any
of these lines the angular momentum remains constant, such that each
line can be labeled by $L$. For ascending field the vortex pattern
moves from $L$ independent vortices at low field to giant vortex
states at high field, whose total angular momentum must add to $L$.
Therefore the present method is able to reproduce the well-known
features of mesoscopic superconductors found by other authors using
different approaches\cite{PhysRevLett.81.2783,PhysRevB.65.104515}.

\begin{table*}
\caption{\label{tab:magmax} Points ($h_L/H_{c2},-4\pi D M_L/H_{c2}$)
are the maximum of the magnetization lines associated to the angular
momentum L curves for the different models.}
\begin{ruledtabular}
\begin{tabular}{|l|l|l|l|l|l|l|}
L&sharp&mesh&sphere&BP2D&smooth&step\\
\hline
0 &(0.38,0.25)  & (0.39,0.25) & (0.43,0.28)    &(0.58,0.44) & (0.39,0.25) &(0.44,0.28)\\
1 & (0.58,0.19) & (0.59,0.20) & (0.66,0.19)   &(0.73,0.39) & (0.59,0.19)& (0.64,0.22)\\
2 &  (0.75,0.14)& (0.75,0.15) & (0.84,0.13)    &(0.86,0.34) & (0.75,0.14) & (0.81,0.17)\\
3 &  (0.91,0.11)& (0.90,0.12) & (0.98,0.090)  &(0.98,0.29) & (0.90,0.10)&(0.98,0.13)\\
4 &  (1.05,0.082)&(1.04,0.089)& (1.14,0.060)    &(1.11,0.24) & (1.04,0.077) &(1.13,0.11) \\
5 &  (1.19,0.060)&(1.18,0.067)& (1.27,0.038)     &(1.23,0.20) & (1.17,0.055)&(1.27,0.083) \\
6 &  (1.32,0.044)&(1.30,0.049)& (1.39,0.022)      &(1.34,0.15) & (1.30,0.038)&(1.41,0.064)\\
7 &  (1.44,0.030)&(1.43,0.034)& (1.50,0.011)    &(1.45,0.12) & (1.41,0.025) &(1.55,0.048)  \\
8 & (1.56,0.019) &(1.54,0.022)& (1.60,0.0034)    &(1.56,0.083) & (1.53,0.015)&(1.67,0.036) \\
9 & (1.67,0.011) &(1.67,0.013)&  -    &(1.66,0.054) & (1.64,0.0072)&(1.80,0.026)\\
10 &  (1.78,0.0046)&(1.77,0.0057) & -     &(1.78,0.029) & (1.74,0.0023)&(1.92,0.016) \\
11 & (1.88,8.8x10$^{-4}$)&(1.86,9.7x10$^{-4}$)&  -  &(1.87,0.0096) & -&(2.03,0.010) \\
12 &  - & - &  -  & -  & -&(2.14,0.0051)  \\
13 &  - & - &  -   & - & -&(2.24,0.0019) \\
\end{tabular}
\end{ruledtabular}
\end{table*}

Table~\ref{tab:hmatch} shows the matching fields $h_{L\,L+1}$ between two nearest angular momentum states for
the six models considered here. Table~\ref{tab:magmax} is useful for model comparison, as it shows the maximum
value of $-4\pi D M_L/H_{c2}$ for each $L$ lines and its corresponding applied field $h_L/H_{c2}$.

\section{Discussion}
\label{sec:dis}

In this section we compare all models to the \textit{sharp} boundary
model for the mesoscopic disk whose properties are shown in
Tables~\ref{tab:hmatch} and ~\ref{tab:magmax}. The effect of the
number of grid points in our calculations can be checked in
Fig.~\ref{mesh_sharp} as the \textit{mesh} model has 8.2 times more
grid points than the \textit{sharp} model. The \textit{mesh} model
has a lower free energy and a higher magnetization than the
\textit{sharp} model, but their values differ by less than one per
cent. Effects due to the grid become only noticeable for
intermediate fields, but not in the single vortex region. The
comparison between \textit{sharp} and \textit{mesh} shows that the
present numerical approach is robust and displays very little
quantitative dependence on the grid. Fig.~\ref{mesh_sharp} also
shows the iso-density three-dimensional plots of four typical vortex
configurations selected to display 1, 2, 3 and 4 vortex states,
respectively. Their corresponding applied field, magnetization and
free energy values can be read from Fig.~\ref{mesh_sharp}.
\begin{figure*}[t]
\includegraphics[width=\linewidth]{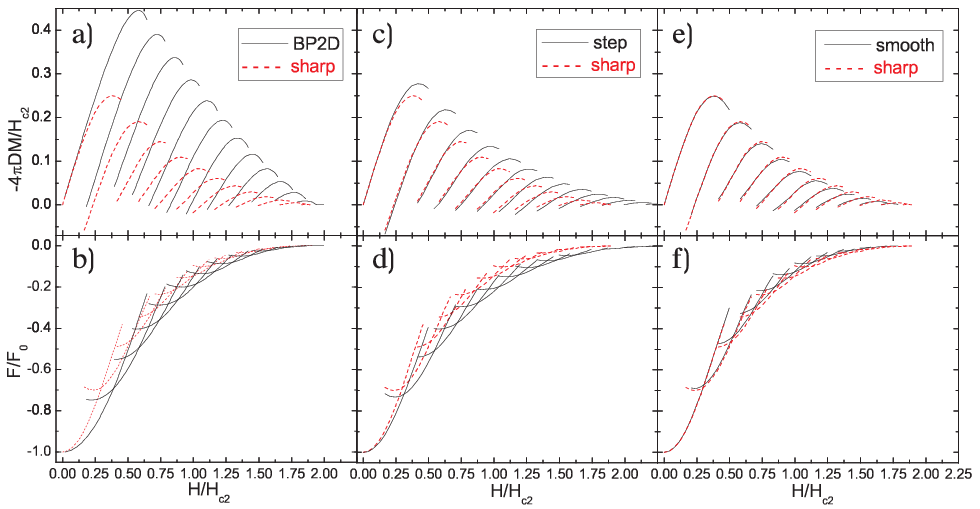}
(Color online)\caption{a) and b): { The two-dimensional
\textit{BP2D} disk and the three-dimensional \textit{sharp}(red)
disk magnetization and free-energy.} c) and d): {The \textit{step}
and the \textit{sharp}(red) disk magnetization and free-energy.} e)
and f): {The \textit{smooth} and the \textit{sharp}(red) disk
magnetization and free-energy.}} \label{fig4}
\end{figure*}

Fig.~\ref{esf_sharp} shows results of the magnetization and free energy for the \textit{sphere} model, and also
for the \textit{sharp} disk model. Three-dimensional iso-density and two-dimensional contour plots are shown for
four selected vortex configurations, whose location is indicated in the free-energy curve. These four vortex
configurations illustrate general features of the vortex lines inside the sphere. A vortex line must reach the
surface perpendicularly in order to avoid a supercurrent component pointing outwards the
surface\cite{j.low.temp.Phys.44,Phys.c.185.1465}. Because of the small volume to surface ratio of this
$R_s=4.0\xi$ sphere, the vortex lines are strongly affected by this surface effect and as a result they are
curved everywhere inside the sphere. The lines are closely packed in the equatorial plane as shown by the
three-dimensional iso-density plots. These plots are drawn at 20\% of the maximum order density and each plot
consists of a single iso-density surface. The north pole part of these plots provide a view of the vortex
behavior at the surface, but the translucent properties of these three-dimensional plots make it difficult to
have the same view at the south-pole. Fig.~\ref{esf_sharp} also shows two-dimensional contour plots associated
to two selected cuts of the sphere, taken at the equatorial (half-plane) and half-way between the north (or
south) pole and the equator planes (quarter-plane). These contour plots contain ten contour regions, shown in
different colors, ranging from maximum density (red) to minimum density (blue). They also show that the vortex
lines are closely packed at the equatorial plane and also that the vortex core is larger near the surface than
inside the sphere. The \textit{sphere} has a stronger magnetic signal as compared to the \textit{disk} for low
fields, but for high fields up to $H_{c3}$  the situation turns and the disk acquires a stronger signal. In fact
the \textit{sphere} only supports 9 vortex states whereas the disk 12 states. As the field increases the vortex
configuration in case of the sphere disappears faster than in the disk, probably due to the existence of vortex
lines of different lengths.

Fig.~\ref{fig4} shows comparative analysis of the free energy and magnetization curves to the \textit{sharp}
model for the \textit{BP2D}, \textit{step}, and \textit{smooth} models. (i)\textit{BP2D-sharp}: Their different
$\kappa$ values yield significantly different magnetization curves. The \textit{BP2D} model has a very strong
diamagnetic signal lower free energy states, because it has a more effective shielding to the applied field.
Nevertheless the models show qualitative similarities. They both have the same number of 12 angular momentum
states, as shown in Table~\ref{tab:hmatch}, and present a fair agreement between matching fields in the high
field region. This is explained by the weakening of the diamagnetic currents for high field that turns the
(\textit{BP2D}) similarly to the \textit{sharp} model.(ii)\textit{step-sharp}: The presence of an intermediate
region at the boundary enlarges surface effects as compared to the \textit{sharp} case. The diamagnetic response
is stronger, and the free energy is lower, in all $L$ lines and in fact it allows for two extra vortex states,
according to Table~\ref{tab:hmatch}. These features are not a consequence of a slight difference in height
between the two models.(iii)\textit{smooth-sharp}: The \textit{smooth} model treats the boundary differently
from \textit{sharp} in case the smooth $\tau$ function decay takes place over a distance larger than the mesh
parameters $a_x$, $a_y$, and $a_z$. This is the case here but we find no substantial change in behavior by using
the \textit{smooth} model. This boundary was extensively used in previous problems of a superconductor with a
periodic array of inclusions\cite{EuroPhysLett.67.446,PhysRevB.66.064519}.

\section{Conclusion}
\label{sec:con} The vortex patterns of truly three-dimensional
mesoscopic superconductors, namely a disk and a sphere, were
analyzed. They were obtained by numerical minimization (Simulated
Annealing) of a modified GL free energy that already incorporates
the boundary conditions. This procedure provides an efficient way to
obtain vortex patterns in mesoscopic superconductors and needs
relatively few grid points. The method is stable under changes of
the grid size, and for a two-dimensional disk it reproduces results
of disk geometry previously studied by other methods
\cite{PhysRevB.65.104515}). We find that slight changes of the
boundary conditions, like the creation of a surface layer, increases
the upper critical field and allows for an increase in the number of
angular momentum states. In case of a mesoscopic sphere we find that
the vortex lines are naturally curved due to strong surface effects
as was recently also found in a wire with a
constriction\cite{europhyslett.74.151}.

\section{Acknowledgement}
\label{Acknowledges} A. R. de C. Romaguera acknowledges support from CNPq (Brazil). M.M. Doria acknowledges
support from CNPq (Brazil), FAPERJ (Brazil), the Instituto do Mil\^enio de Nanotecnologia (Brazil) and BOF/UA
(Belgium). F. M. Peeters acknowledges support from the Flemish Science Foundation (FWO-Vl), the Belgian Science
Policy (IUAP) and the ESF-AQDJJ network.
\bibliographystyle{apsrev}


\bibliography{meso}

\begin{thebibliography}{28}
\expandafter\ifx\csname natexlab\endcsname\relax\def\natexlab#1{#1}\fi
\expandafter\ifx\csname bibnamefont\endcsname\relax
  \def\bibnamefont#1{#1}\fi
\expandafter\ifx\csname bibfnamefont\endcsname\relax
  \def\bibfnamefont#1{#1}\fi
\expandafter\ifx\csname citenamefont\endcsname\relax
  \def\citenamefont#1{#1}\fi
\expandafter\ifx\csname url\endcsname\relax
  \def\url#1{\texttt{#1}}\fi
\expandafter\ifx\csname urlprefix\endcsname\relax\def\urlprefix{URL }\fi
\providecommand{\bibinfo}[2]{#2}
\providecommand{\eprint}[2][]{\url{#2}}

\bibitem[{\citenamefont{Schweigert et~al.}(1998)\citenamefont{Schweigert,
  Peeters, and Deo}}]{PhysRevLett.81.2783}
\bibinfo{author}{\bibfnamefont{V.~A.} \bibnamefont{Schweigert}},
  \bibinfo{author}{\bibfnamefont{F.~M.} \bibnamefont{Peeters}},
  \bibnamefont{and} \bibinfo{author}{\bibfnamefont{P.~S.} \bibnamefont{Deo}},
  \bibinfo{journal}{Phys. Rev. Lett.} \textbf{\bibinfo{volume}{81}},
  \bibinfo{pages}{2783} (\bibinfo{year}{1998}).

\bibitem[{\citenamefont{Schweigert and Peeters}(1998)}]{PhysRevB.57.13817}
\bibinfo{author}{\bibfnamefont{V.~A.} \bibnamefont{Schweigert}}
  \bibnamefont{and} \bibinfo{author}{\bibfnamefont{F.~M.}
  \bibnamefont{Peeters}}, \bibinfo{journal}{Phys. Rev. B}
  \textbf{\bibinfo{volume}{57}}, \bibinfo{pages}{13817} (\bibinfo{year}{1998}).

\bibitem[{\citenamefont{Geim et~al.}(1997)\citenamefont{Geim, Grigorieva,
  Dubonos, Lok, Maan, Filippov, and Peeters}}]{Nature.390.259}
\bibinfo{author}{\bibfnamefont{A.~K.} \bibnamefont{Geim}},
  \bibinfo{author}{\bibfnamefont{I.~V.} \bibnamefont{Grigorieva}},
  \bibinfo{author}{\bibfnamefont{S.~V.} \bibnamefont{Dubonos}},
  \bibinfo{author}{\bibfnamefont{J.~G.~S.} \bibnamefont{Lok}},
  \bibinfo{author}{\bibfnamefont{J.~C.} \bibnamefont{Maan}},
  \bibinfo{author}{\bibfnamefont{A.~E.} \bibnamefont{Filippov}},
  \bibnamefont{and} \bibinfo{author}{\bibfnamefont{F.~M.}
  \bibnamefont{Peeters}}, \bibinfo{journal}{Nature (London)}
  \textbf{\bibinfo{volume}{390}}, \bibinfo{pages}{259} (\bibinfo{year}{1997}).

\bibitem[{\citenamefont{Kanda et~al.}(2004)\citenamefont{Kanda, Baelus,
  Peeters, Kadowaki, and Ootuka}}]{PhysRevLett.93.257002}
\bibinfo{author}{\bibfnamefont{A.}~\bibnamefont{Kanda}},
  \bibinfo{author}{\bibfnamefont{B.~J.} \bibnamefont{Baelus}},
  \bibinfo{author}{\bibfnamefont{F.~M.} \bibnamefont{Peeters}},
  \bibinfo{author}{\bibfnamefont{K.}~\bibnamefont{Kadowaki}}, \bibnamefont{and}
  \bibinfo{author}{\bibfnamefont{Y.}~\bibnamefont{Ootuka}},
  \bibinfo{journal}{Phys. Rev. Lett.} \textbf{\bibinfo{volume}{93}},
  \bibinfo{pages}{257002} (\bibinfo{year}{2004}).

\bibitem[{\citenamefont{Doria and de~C.~Romaguera}(2004)}]{EuroPhysLett.67.446}
\bibinfo{author}{\bibfnamefont{M.~M.} \bibnamefont{Doria}} \bibnamefont{and}
  \bibinfo{author}{\bibfnamefont{A.~R.} \bibnamefont{de~C.~Romaguera}},
  \bibinfo{journal}{Europhys. Lett.} \textbf{\bibinfo{volume}{67}},
  \bibinfo{pages}{446} (\bibinfo{year}{2004}), \eprint{cond-mat/0407599}.

\bibitem[{\citenamefont{Doria and Zebende}(2002)}]{PhysRevB.66.064519}
\bibinfo{author}{\bibfnamefont{M.~M.} \bibnamefont{Doria}} \bibnamefont{and}
  \bibinfo{author}{\bibfnamefont{G.~F.} \bibnamefont{Zebende}},
  \bibinfo{journal}{Phys. Rev. B} \textbf{\bibinfo{volume}{66}},
  \bibinfo{pages}{064519} (\bibinfo{year}{2002}).

\bibitem[{\citenamefont{Mkrtchyan and Schmidt}(1972)}]{SovietPhys.JETP.34}
\bibinfo{author}{\bibfnamefont{G.~S.} \bibnamefont{Mkrtchyan}}
  \bibnamefont{and} \bibinfo{author}{\bibfnamefont{V.~V.}
  \bibnamefont{Schmidt}}, \bibinfo{journal}{Soviet Phys. JETP}
  \textbf{\bibinfo{volume}{34}}, \bibinfo{pages}{195} (\bibinfo{year}{1972}).

\bibitem[{\citenamefont{Buzdin}(1993)}]{PhysRevB.47.11416}
\bibinfo{author}{\bibfnamefont{A.~I.} \bibnamefont{Buzdin}},
  \bibinfo{journal}{Phys. Rev. B} \textbf{\bibinfo{volume}{47}},
  \bibinfo{pages}{11416} (\bibinfo{year}{1993}).

\bibitem[{\citenamefont{Doria and de~Andrade}(1999)}]{PhysRevB.60.13164}
\bibinfo{author}{\bibfnamefont{M.~M.} \bibnamefont{Doria}} \bibnamefont{and}
  \bibinfo{author}{\bibfnamefont{S.~C.~B.} \bibnamefont{de~Andrade}},
  \bibinfo{journal}{Phys. Rev. B} \textbf{\bibinfo{volume}{60}},
  \bibinfo{pages}{13164} (\bibinfo{year}{1999}).

\bibitem[{\citenamefont{Berdiyorov et~al.}(2006)\citenamefont{Berdiyorov,
  Milo\ifmmode \check{s}\else \v{s}\fi{}evi\ifmmode~\acute{c}\else \'{c}\fi{},
  and Peeters}}]{PhysRevLett.96.207001}
\bibinfo{author}{\bibfnamefont{G.~R.} \bibnamefont{Berdiyorov}},
  \bibinfo{author}{\bibfnamefont{M.~V.} \bibnamefont{Milo\ifmmode
  \check{s}\else \v{s}\fi{}evi\ifmmode~\acute{c}\else \'{c}\fi{}}},
  \bibnamefont{and} \bibinfo{author}{\bibfnamefont{F.~M.}
  \bibnamefont{Peeters}}, \bibinfo{journal}{Phys. Rev. Lett.}
  \textbf{\bibinfo{volume}{96}}, \bibinfo{pages}{207001}
  (\bibinfo{year}{2006}).

\bibitem[{\citenamefont{{Raedts} et~al.}(2004)\citenamefont{{Raedts},
  {Silhanek}, {van Bael}, and {Moshchalkov}}}]{PhysRevB.70.24509}
\bibinfo{author}{\bibfnamefont{S.}~\bibnamefont{{Raedts}}},
  \bibinfo{author}{\bibfnamefont{A.~V.} \bibnamefont{{Silhanek}}},
  \bibinfo{author}{\bibfnamefont{M.~J.} \bibnamefont{{van Bael}}},
  \bibnamefont{and} \bibinfo{author}{\bibfnamefont{V.~V.}
  \bibnamefont{{Moshchalkov}}}, \bibinfo{journal}{Phys. Rev. B}
  \textbf{\bibinfo{volume}{70}}, \bibinfo{pages}{24509} (\bibinfo{year}{2004}).

\bibitem[{\citenamefont{Bezryadin et~al.}(1995)\citenamefont{Bezryadin, Buzdin,
  and Pannetier}}]{PhysRevB.51.3718}
\bibinfo{author}{\bibfnamefont{A.}~\bibnamefont{Bezryadin}},
  \bibinfo{author}{\bibfnamefont{A.}~\bibnamefont{Buzdin}}, \bibnamefont{and}
  \bibinfo{author}{\bibfnamefont{B.}~\bibnamefont{Pannetier}},
  \bibinfo{journal}{Phys. Rev. B} \textbf{\bibinfo{volume}{51}},
  \bibinfo{pages}{3718} (\bibinfo{year}{1995}).

\bibitem[{\citenamefont{Bezryadin et~al.}(1994)\citenamefont{Bezryadin, Buzdin,
  and Pannetier}}]{PhysLettA.195.373}
\bibinfo{author}{\bibfnamefont{A.}~\bibnamefont{Bezryadin}},
  \bibinfo{author}{\bibfnamefont{A.}~\bibnamefont{Buzdin}}, \bibnamefont{and}
  \bibinfo{author}{\bibfnamefont{B.}~\bibnamefont{Pannetier}},
  \bibinfo{journal}{Phys. Lett. A} \textbf{\bibinfo{volume}{195}},
  \bibinfo{pages}{373} (\bibinfo{year}{1994}).

\bibitem[{\citenamefont{Moshchalkov et~al.}(1996)\citenamefont{Moshchalkov,
  Baert, Metlushko, Rosseel, Van~Bael, Temst, Jonckheere, and
  Bruynseraede}}]{PhysRevB.54.7385}
\bibinfo{author}{\bibfnamefont{V.~V.} \bibnamefont{Moshchalkov}},
  \bibinfo{author}{\bibfnamefont{M.}~\bibnamefont{Baert}},
  \bibinfo{author}{\bibfnamefont{V.~V.} \bibnamefont{Metlushko}},
  \bibinfo{author}{\bibfnamefont{E.}~\bibnamefont{Rosseel}},
  \bibinfo{author}{\bibfnamefont{M.~J.} \bibnamefont{Van~Bael}},
  \bibinfo{author}{\bibfnamefont{K.}~\bibnamefont{Temst}},
  \bibinfo{author}{\bibfnamefont{R.}~\bibnamefont{Jonckheere}},
  \bibnamefont{and}
  \bibinfo{author}{\bibfnamefont{Y.}~\bibnamefont{Bruynseraede}},
  \bibinfo{journal}{Phys. Rev. B} \textbf{\bibinfo{volume}{54}},
  \bibinfo{pages}{7385} (\bibinfo{year}{1996}).

\bibitem[{\citenamefont{{Silhanek} et~al.}(2004)\citenamefont{{Silhanek},
  {Raedts}, {van Bael}, and {Moshchalkov}}}]{PhysRevB.70.54515}
\bibinfo{author}{\bibfnamefont{A.~V.} \bibnamefont{{Silhanek}}},
  \bibinfo{author}{\bibfnamefont{S.}~\bibnamefont{{Raedts}}},
  \bibinfo{author}{\bibfnamefont{M.~J.} \bibnamefont{{van Bael}}},
  \bibnamefont{and} \bibinfo{author}{\bibfnamefont{V.~V.}
  \bibnamefont{{Moshchalkov}}}, \bibinfo{journal}{Phys. Rev. B}
  \textbf{\bibinfo{volume}{70}}, \bibinfo{pages}{054515}
  (\bibinfo{year}{2004}).

\bibitem[{\citenamefont{Murakami et~al.}(1996)\citenamefont{Murakami, Sakai,
  Higuchi, and Yoo}}]{sup.sci.tech.9.1015}
\bibinfo{author}{\bibfnamefont{M.}~\bibnamefont{Murakami}},
  \bibinfo{author}{\bibfnamefont{N.}~\bibnamefont{Sakai}},
  \bibinfo{author}{\bibfnamefont{T.}~\bibnamefont{Higuchi}}, \bibnamefont{and}
  \bibinfo{author}{\bibfnamefont{S.~I.} \bibnamefont{Yoo}},
  \bibinfo{journal}{Supercond. Sci. Technol.} \textbf{\bibinfo{volume}{9}},
  \bibinfo{pages}{1015} (\bibinfo{year}{1996}).

\bibitem[{\citenamefont{Shi et~al.}(2005)\citenamefont{Shi, Babu, and
  Cardwell}}]{sup.sci.tech.18.l13}
\bibinfo{author}{\bibfnamefont{Y.}~\bibnamefont{Shi}},
  \bibinfo{author}{\bibfnamefont{N.~H.} \bibnamefont{Babu}}, \bibnamefont{and}
  \bibinfo{author}{\bibfnamefont{D.~A.} \bibnamefont{Cardwell}},
  \bibinfo{journal}{Supercond. Sci. Technol.} \textbf{\bibinfo{volume}{18}},
  \bibinfo{pages}{L13} (\bibinfo{year}{2005}).

\bibitem[{\citenamefont{Du}(2005)}]{j.math.phts.46.095109}
\bibinfo{author}{\bibfnamefont{Q.}~\bibnamefont{Du}}, \bibinfo{journal}{J.
  Math. Phys.} \textbf{\bibinfo{volume}{46}}, \bibinfo{pages}{095109}
  (\bibinfo{year}{2005}).

\bibitem[{\citenamefont{Elmurodov et~al.}(2006)\citenamefont{Elmurodov,
  Vodolazov, and Peeters}}]{europhyslett.74.151}
\bibinfo{author}{\bibfnamefont{A.~K.} \bibnamefont{Elmurodov}},
  \bibinfo{author}{\bibfnamefont{D.~Y.} \bibnamefont{Vodolazov}},
  \bibnamefont{and} \bibinfo{author}{\bibfnamefont{F.~M.}
  \bibnamefont{Peeters}}, \bibinfo{journal}{Europhys. Lett.}
  \textbf{\bibinfo{volume}{74}}, \bibinfo{pages}{151} (\bibinfo{year}{2006}).

\bibitem[{\citenamefont{Berdiyorov et~al.}(2004)\citenamefont{Berdiyorov,
  Milosevic, Baelus, and Peeters}}]{PhysRevB.70.024508}
\bibinfo{author}{\bibfnamefont{G.}~\bibnamefont{Berdiyorov}},
  \bibinfo{author}{\bibfnamefont{M.}~\bibnamefont{Milosevic}},
  \bibinfo{author}{\bibfnamefont{B.~J.} \bibnamefont{Baelus}},
  \bibnamefont{and} \bibinfo{author}{\bibfnamefont{F.~M.}
  \bibnamefont{Peeters}}, \bibinfo{journal}{Phys. Rev. B}
  \textbf{\bibinfo{volume}{70}}, \bibinfo{pages}{024508}
  (\bibinfo{year}{2004}).

\bibitem[{\citenamefont{Baelus and Peeters}(2002)}]{PhysRevB.65.104515}
\bibinfo{author}{\bibfnamefont{B.~J.} \bibnamefont{Baelus}} \bibnamefont{and}
  \bibinfo{author}{\bibfnamefont{F.~M.} \bibnamefont{Peeters}},
  \bibinfo{journal}{Phys. Rev. B} \textbf{\bibinfo{volume}{65}},
  \bibinfo{pages}{104515} (\bibinfo{year}{2002}).

\bibitem[{\citenamefont{Saint-James and {de Gennes}}(1963)}]{Phys.Lett.7.306}
\bibinfo{author}{\bibfnamefont{D.}~\bibnamefont{Saint-James}} \bibnamefont{and}
  \bibinfo{author}{\bibfnamefont{P.~G.} \bibnamefont{{de Gennes}}},
  \bibinfo{journal}{Phys. Lett.} \textbf{\bibinfo{volume}{7}},
  \bibinfo{pages}{306} (\bibinfo{year}{1963}).

\bibitem[{\citenamefont{Doria and de~C.~Romaguera}(2005)}]{Brazilian.J.Phys.35}
\bibinfo{author}{\bibfnamefont{M.~M.} \bibnamefont{Doria}} \bibnamefont{and}
  \bibinfo{author}{\bibfnamefont{A.~R.} \bibnamefont{de~C.~Romaguera}},
  \bibinfo{journal}{Brazilian J. Phys.} \textbf{\bibinfo{volume}{35}},
  \bibinfo{pages}{157} (\bibinfo{year}{2005}), \eprint{cond-mat/0411046}.

\bibitem[{\citenamefont{de~C.~Romaguera and Doria}(2004)}]{Eur.Phys.J.B.42}
\bibinfo{author}{\bibfnamefont{A.~R.} \bibnamefont{de~C.~Romaguera}}
  \bibnamefont{and} \bibinfo{author}{\bibfnamefont{M.~M.} \bibnamefont{Doria}},
  \bibinfo{journal}{Eur. Phys. J. B} \textbf{\bibinfo{volume}{42}},
  \bibinfo{pages}{3} (\bibinfo{year}{2004}).

\bibitem[{\citenamefont{Doria et~al.}(1990)\citenamefont{Doria, Gubernatis, and
  Rainer}}]{PhysRevB.41.6335}
\bibinfo{author}{\bibfnamefont{M.~M.} \bibnamefont{Doria}},
  \bibinfo{author}{\bibfnamefont{J.~E.} \bibnamefont{Gubernatis}},
  \bibnamefont{and} \bibinfo{author}{\bibfnamefont{D.}~\bibnamefont{Rainer}},
  \bibinfo{journal}{Phys. Rev. B} \textbf{\bibinfo{volume}{41}},
  \bibinfo{pages}{6335} (\bibinfo{year}{1990}).

\bibitem[{\citenamefont{Doria et~al.}(1989)\citenamefont{Doria, Gubernatis, and
  Rainer}}]{PhysRevB.39.9573}
\bibinfo{author}{\bibfnamefont{M.~M.} \bibnamefont{Doria}},
  \bibinfo{author}{\bibfnamefont{J.~E.} \bibnamefont{Gubernatis}},
  \bibnamefont{and} \bibinfo{author}{\bibfnamefont{D.}~\bibnamefont{Rainer}},
  \bibinfo{journal}{Phys. Rev. B} \textbf{\bibinfo{volume}{39}},
  \bibinfo{pages}{9573} (\bibinfo{year}{1989}).

\bibitem[{\citenamefont{Brandt}(1981)}]{j.low.temp.Phys.44}
\bibinfo{author}{\bibfnamefont{E.~H.} \bibnamefont{Brandt}},
  \bibinfo{journal}{J. Low Temp. Phys.} \textbf{\bibinfo{volume}{44}},
  \bibinfo{pages}{33} (\bibinfo{year}{1981}).

\bibitem[{\citenamefont{{Buisson} et~al.}(1991)\citenamefont{{Buisson},
  {Carneiro}, and {Doria}}}]{Phys.c.185.1465}
\bibinfo{author}{\bibfnamefont{O.}~\bibnamefont{{Buisson}}},
  \bibinfo{author}{\bibfnamefont{G.}~\bibnamefont{{Carneiro}}},
  \bibnamefont{and} \bibinfo{author}{\bibfnamefont{M.}~\bibnamefont{{Doria}}},
  \bibinfo{journal}{Physica C} \textbf{\bibinfo{volume}{185}},
  \bibinfo{pages}{1465} (\bibinfo{year}{1991}).

\end{thebibliography}
\end{document}